\documentclass{vgtc}                          
\usepackage[belowskip=0pt,aboveskip=5pt]{caption}
\usepackage{enumitem}

\ifpdf
  \pdfoutput=1\relax                   
  \usepackage{graphicx}                
  \DeclareGraphicsExtensions{.pdf,.png,.jpg,.jpeg} 
\else
  \usepackage{graphicx}                
  \DeclareGraphicsExtensions{.eps}     
\fi%

\graphicspath{{figures/}{pictures/}{images/}{./}} 

\usepackage{microtype}                 
\PassOptionsToPackage{warn}{textcomp}  
\usepackage{textcomp}                  
\usepackage{mathptmx}                  
\usepackage{times}                     
\usepackage{cite}                      
\usepackage{tabu}                      
\usepackage{booktabs}                  

\onlineid{0}

\vgtccategory{Research}

\vgtcinsertpkg

\title{Creating Informal Learning and First Responder Training XR Experiences with the ImmersiveDeck}

\author{
Christian Schönauer\thanks{e-mail: christian.schoenauer@tuwien.ac.at}\\
Hannes Kaufmann
\\ %
        \scriptsize TU Wien %
\and 
Maria Roussou\thanks{e-mail: mroussou@di.uoa.gr}\\ %
     \scriptsize National and Kapodistrian \\\scriptsize University of Athens %
\and 
Julien Rüggeberg\thanks{e-mail: julien@illusion-walk.com}\\
Jim Rüggeberg
\\ %
\scriptsize Illusion Walk KG 
\and 
Lucas Katsikaris\thanks{e-mail: lucas@boltvirtual.gr}\\
Sakis Rogkas
\\ %
\scriptsize Bolt Virtual
\and 
Dimitris Christopoulos\thanks{e-mail: christop@fhw.gr}
\\ %
\scriptsize Foundation of the Hellenic World
\vspace{-1em}
}

\teaser{
  \hspace*{-0.05\textwidth} 
   \includegraphics[width=1.1\textwidth]{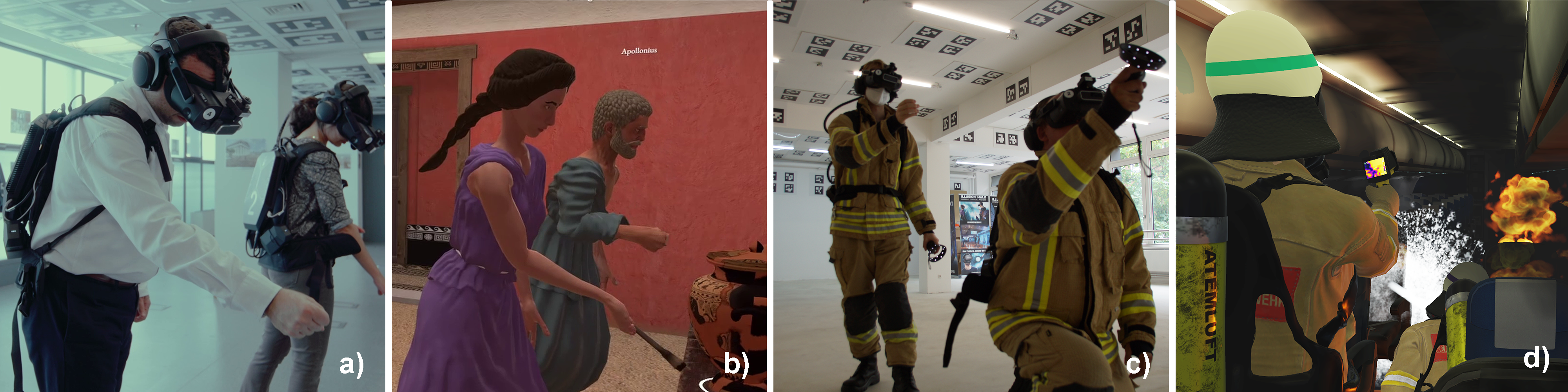}
 \caption{a) Users of the ImmersiveDeck platform in the b) informal learning experience; c) Firefighters with equipment in the\\d) Aircraft Rescue and Fire Fighting XR Training.}
 \label{fig:Teaser}
}

\abstract{
In recent years eXtended Reality (XR) technologies have matured and have become affordable, yet creating XR experiences for training and learning in many cases is still a time-consuming and costly process, hindering widespread adoption. One factor driving effort is that content and features commonly required by many applications get re-implemented for each experience, instead of sharing and reusing these resources by means of a common platform.
In this paper we present two XR experiences in the context of informal learning and first responder training along with the shared platform they have been created with and the creation process.
Furthermore, we have technically evaluated relevant parts of the platform for feasibility of use with experience requirements and confirmed applicability.
Finally, we present an informal expert evaluation of the content creation process’s user experience for the informal learning experience along with guidelines derived from the findings.

} 

\CCScatlist{
  \CCScatTwelve{Human-centered computing}{Visu\-al\-iza\-tion};
  \CCScatTwelve{Human-centered computing}{Visu\-al\-iza\-tion}{Visualization design and evaluation methods}{}
}

\begin{document}


\firstsection{Introduction}

\maketitle

There is widespread scientific consent that (immersive) ``VR is an extremely promising tool for the enhancement of learning, education, and training'' \cite{slater2016enhancing}. However, despite technological advancements in screen, optics and tracking technology leading to affordable and accessible head-mounted displays (HMDs), the adoption of these technologies in learning, training, professional but also recreational activities is still hampered by several shortcomings.

Firstly, content development is still the most expensive part of a location-based XR experience. Most platform providers allow only content out of their studio or produce it only in cooperation with a specific customer. The profitability of the content relies strongly on economies of scale. For high volume experiences in the entertainment sector one production can be run in various locations, but for educational, training or cultural experiences the high cost per user is still a major problem for location-based XR. One factor driving the cost of content creation is that experiences are often created as closed applications, for which many features like interaction, multi-user management, collaboration etc. have to be (re-)implemented with each new experience.

Secondly, the digital, virtual world in most immersive setups is still well separated from the physical world, making the connection between the two distinct. This is an issue especially where such connection is important for adding value, e.g. in simulations where the mapping between the virtual and physical world enhances the fidelity and believability of the simulation, thus significantly increasing the sense of presence and the effectiveness of the situated training or learning scenario \cite{Yannier2013Tangible}.

Furthermore, the stimulation of senses, other than the dominant visual and auditory senses, has been underexplored in VR as well as in AR and MR applications. For example, touch in XR is lacking and its integration with immersive environments, to date, has been under-researched. The significance of the tactile sense for the human experience is well documented \cite{gallace2014touch} while multisensorial augmentation (e.g. olfactory stimuli) has the potential to improve validity and enhance the users’ sense of presence in the virtual environment \cite{kent2016should}.

Our contribution to the field of XR includes:
\vspace{-1em}
\begin{itemize}[noitemsep]
    \item Creation of two XR experiences in the context of informal learning and first responder training with a shared platform.
    \item Development/extension of said platform to support all features required in the respective learning and training domains.
    \item Technical evaluation of relevant platform features.
    \item Preliminary informal evaluation of the content creation process's user experience with experts and derived guidelines.
\end{itemize}

\section{Related Work}

The use of immersive XR in the training of first responders has been an active research area for over two decades~\cite{Stansfield1999} and has been emphasized in recent years due to the progress in XR technology \cite{mossel2021immersive,Engelbrecht_2019_SWOTFirefighting,Hsu2013}. 
Advantages such as safe, cost-efficient, flexible, and repeatable training and review possibilities have led to the development of various scientific and commercial systems \cite{Hsu2013}, especially in the firefighting domain \cite{Engelbrecht_2019_SWOTFirefighting,mossel2021immersive}.
The work of Koutitas et al. with an immersive VR and AR system allowing training in an ambulance ~\cite{Koutitas2019,Koutitas2020} even shows that VR and AR training can outperform infrequent traditional training and improve accuracy. However, the system is limited to a specific restricted environment and specific tasks. 
The commercial systems Advanced Disaster Management Simulator (ADMS)~\cite{ETC-Simulation0000Web} and XVR - Virtual Reality Training Software for Safety and Security~\cite{XVR0000Web} offer non-immersive virtual environments for training incident command and disaster management. While XVR also offers an immersive VR option, it uses a tethered setup restricting the users' navigation by real walking and employs controllers for interaction. Finally, to our knowledge no system exists that allows the training of firefighting in an airplane incident in the flexibility and depth that our experience provides.
 
The benefits of immersive XR in learning have been highlighted many times \cite{perry2017engaging,roussou2004Learning,meccawy2022Creating,roussou2021requirements}. 
In informal learning, especially the domain of cultural heritage, in XR we can (re)create scenarios for a (cultural) learning scenario that would be very costly (buildings, actors, etc.) or outright impossible (landscapes that have been obstructed by modern houses etc.). Furthermore, the process of exploring, discovering, and identifying afforded by immersive VR environments, fosters inquiry that can lead to re-enactment, perspective-taking, and knowledge of the cultural, sociological, and even political importance of objects, ideas, or sites. As much as the term “empathy machine” has been overused, one of the VR medium’s central tenets is indeed this notion that it creates empathy, which in turn, can lead to more radical impacts including the facilitation of attitudinal and value change, attention restoration, therapeutic change and personal transformation \cite{perry2017engaging,roussou2021requirements}. 

For content and experience creation in learning, education, and training a number of general purpose as well as dedicated development tools and platforms exist as listed in \cite{meccawy2022Creating}. However, most of these can produce only applications for desktop and smartphones/tablets or immersive platforms like Enduvo \cite{Enduvo0000Web}, and are limited to standard small-scale VR setups and respective interactions.

In recent years, development of XR platforms has shifted strongly from the academic to the commercial sector. Many of these platforms are targeting location-based entertainment, present proprietary content and are closed to external developers.



Commercial XR platforms like ADMS and XVR provide editors to create first responder training scenarios from ready-made content, but lack flexibility in integration of custom content and are limited to their relatively narrow field of training applications. Furthermore, they are expensive, especially when using proprietary hardware like \cite{Re-lion0000Web}. Most commercial platforms for location-based VR like The VOID \cite{TheVoid0000Web}, Zero Latency \cite{ZeroLatency0000Web} and Sandbox VR \cite{SandboxVR0000Web} are limited to in-house created and/or entertainment applications.
Others like PolygonVR \cite{PolygonVR0000Web} and VIROO \cite{VIROO0000Web} are marketed as open for serious applications also, but use expensive tracking solutions like VIROO's Inmerso technology requiring a dense grid of active panels on the ceiling or are limited room-scale and in features like Vree SDK \cite{Vree0000Web}.

\section{The ImmersiveDeck}
\begin{figure*}[tb]
 \centering 
 \includegraphics[width=\textwidth]{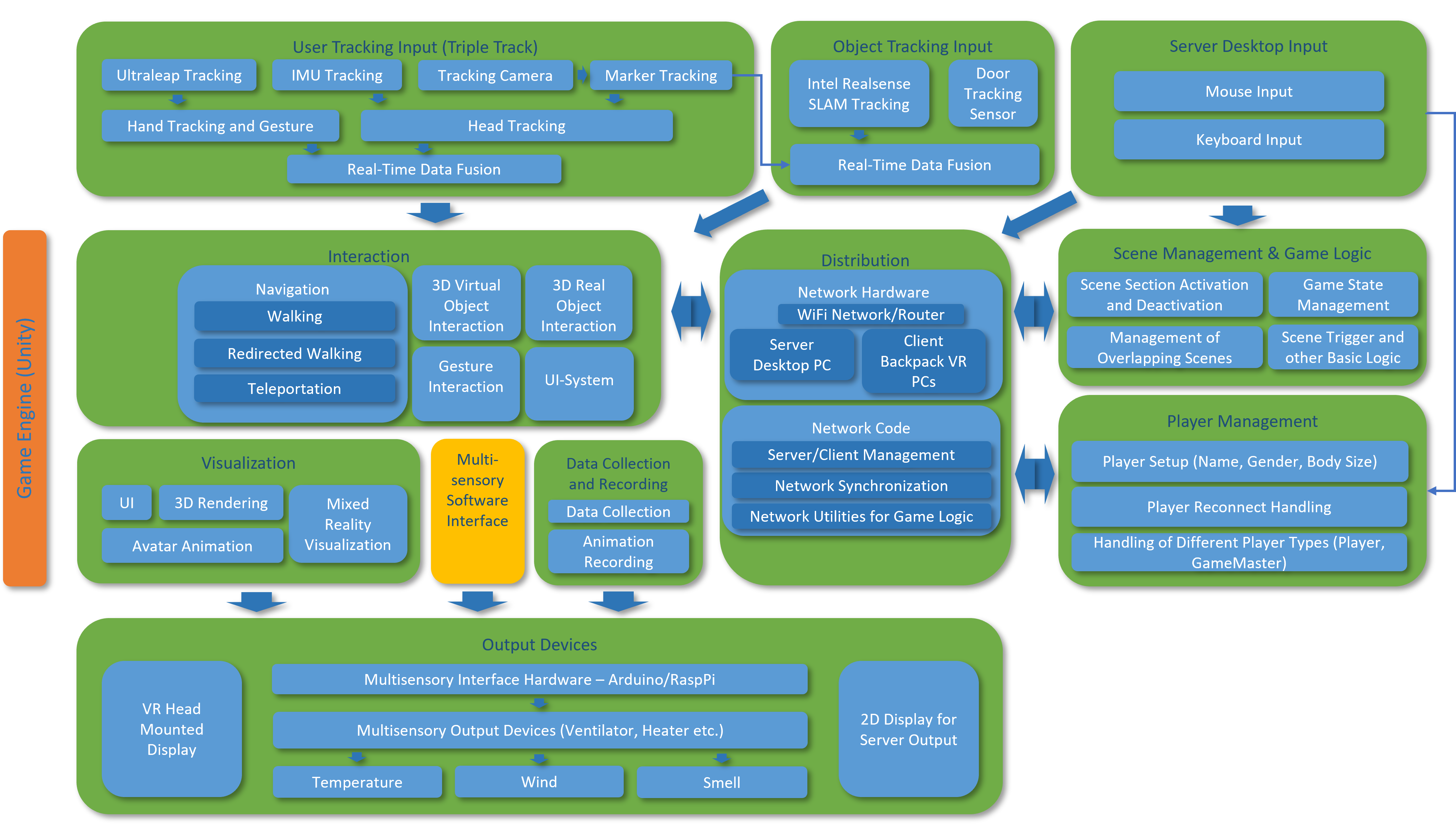}
 \caption{Overview of the ImmersiveDeck platform design and data flow}
 \label{fig:ImmersiveDeck_Platform}
\end{figure*}

The ImmersiveDeck is a low-cost multi-user XR hard- and software platform. It has been first introduced in \cite{podkosova2016ImmersiveDeck} and provided the basis for scientific and commercial XR experiences \cite{schonauer2020physical}. However, recently we enhanced many technical and operational aspects of the platform, which in this context makes a short reintroduction necessary. \autoref{fig:ImmersiveDeck_Platform} shows the design and data flow of the ImmersiveDeck.

The Immersive Deck is a large-scale multi-user XR platform that includes positional tracking, even from room to room, for up to 10 people. Each user wears a Head Mounted Display (HMD) - the HP Reverb G2 or Oculus Rift CV1 - which is tethered to a VR backpack PC (HP Z VR backpack G2), so rendering of the virtual world in the HMD is performed locally on each user to minimize latency. Optical hand and finger tracking is performed by a stereo-camera (Ultraleap) that is attached to the HMD and thus allows the tracking of hand movements for interaction. This setup enables free-roam navigation (simply by walking), by adding absolute optical head tracking using markers, and natural hand interaction. The whole hardware setup is shown in \autoref{fig:Teaser} a)/c) and a close-up of the user-tracking components in \autoref{fig:Athenian_House} c). Software-wise the platform is largely based on the Unity game engine and editor, while software for user- and object-tracking is encapsulated as a separate process and assembly implemented in C++ and C\#.

The ImmersiveDeck is flexible (in terms of applicability for different experiences) and scalable (both in physical training area and number of users). The application focus reported in this paper is in informal learning and (first responder) training, but in principle the platform is open for other areas as well.

\subsection{Content Creation and Integration}
The ImmersiveDeck facilitates the easy creation and integration of content, providing an encapsulated API for anyone to incorporate their content into the platform. The goal is for the ImmersiveDeck solution to lower significantly the entry threshold for content providers to develop content for a multi-user multi-sensory XR platform, i.e. they should not need to be large development companies with dozens of programmers and designers to produce content for such a platform. The final goal is the “democratisation” of XR, a scalable solution for which everybody can freely and openly produce content.
%

An important tool in the content creation architecture of the ImmersiveDeck is the Project-Wizard. With this tool it is possible for content creators to quickly and easily create their custom-configured project from a template (bootstrap-project). The Project-Wizard sets up a minimal project already containing basic functionality like configuration, player management, scene management, tracking and networking. For simple experiences, the scene needs only to be enriched with visual content and can be compiled into a working binary providing all functionality for multi-user interaction.

For more complex experiences the ImmersiveDeck provides a number of features to easily extend content and functionalities. A number of specialized tools accessible via UI-elements in the Unity-Editor make work more intuitive for content creators. Standard configuration for interactions with 3D objects (being grabbable by the user or attached to the user, buttons, touchpad interaction, opening of drawers, doors etc.) are easily applied using the appropriate scripts provided by the platform.

\begin{figure}[tb]
 \centering 
 \includegraphics[width=\columnwidth]{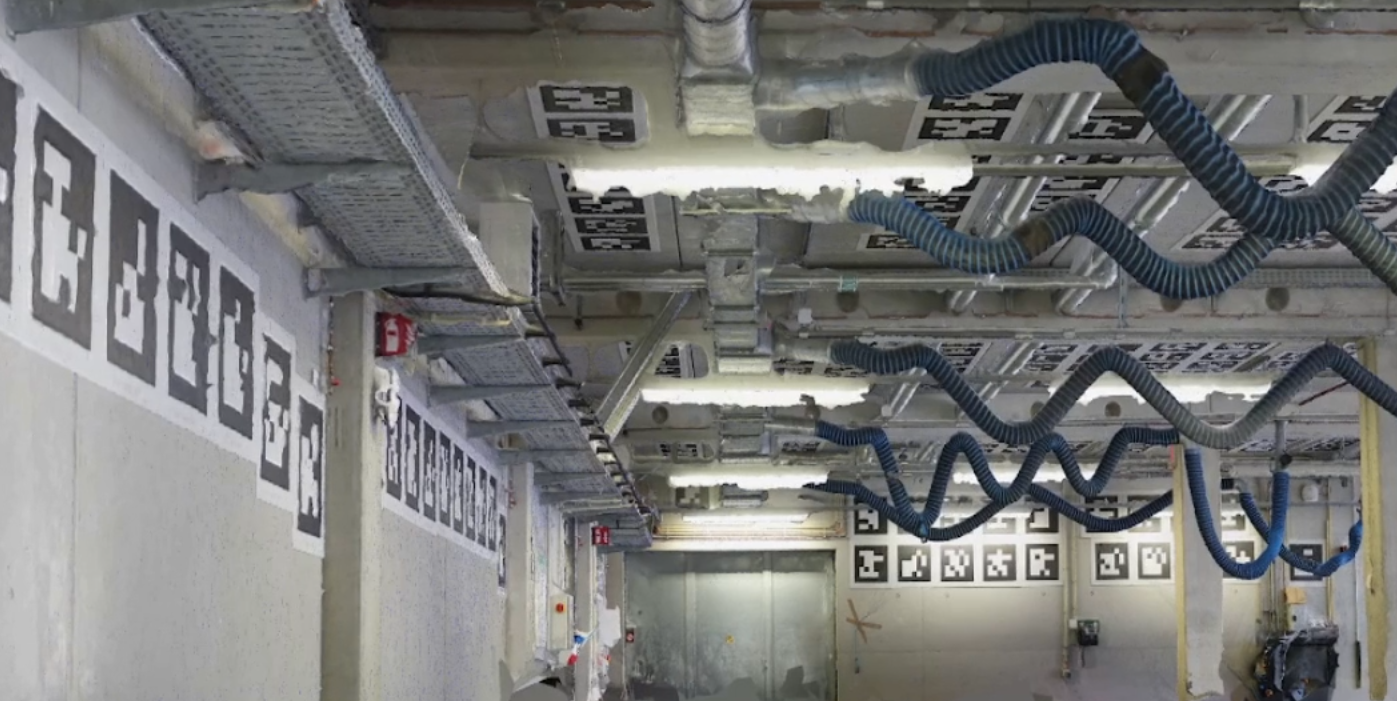}
 \caption{Detailed reconstruction of a fire station garage, created in our semi-automatic approach.}
 \label{fig:Reconstr_Firestation_Garage}
\end{figure}

The ImmersiveDeck supports creation of new user avatars through scripts and tools integrated with the UI of the Unity Editor. Customization of these avatars (size, proportion, color, face-model from user photos) can be easily done by the administrator from the server-UI. 
In addition, the platform provides functionality for scene management and game state management, aiding temporal and spatial organization of the experience. Firstly, this allows to break up complex experiences into multiple scene(-sections), switching between these, and implementation of multiple virtual floors in the same physical space. Secondly, story-lines or sequential steps of an experience can be broken up into synchronized logic states and sequences thereof. This especially helps in situations, where the system administrator wants to advance the experience for all users simultaneously in a distributed environment or a user needs to (re)connect to the system later at a specific state. Also it allows to reset the experience quickly for new users or in case of operational issues.
The ImmersiveDeck, furthermore, provides a feature-rich but low-cost solution for tracking the physical area and mapping the virtual world on to the built environment, including the architecture, objects, machinery, and any artefact related to the scenario of each application, as described in the next subsection. The result of such a reconstruction can be seen in \autoref{fig:Reconstr_Firestation_Garage}, where we mapped a fire station garage.

\subsection{Tracking and Mapping}
\label{subsec:ID:TrackingAndMapping}
The design of the user tracking module is based on the Triple Track® technology, which fuses multiple tracking input sources. Optical and inertial head-tracking are combined with hand-/body-tracking to provide an immersive and interactive experience. The user tracking module adopts two main principles. Firstly, the tracking input is easily usable and configurable for content creators from within the Unity editor to facilitate interaction design. Secondly, implementation details of tracking devices, data handling/fusion algorithms etc. are decoupled from the rendering engine and hidden from said users. This way, we can keep interfaces simple.
To achieve accurate registration between real and virtual content, we implemented a marker tracking solution based on the markers and algorithms of \cite{Shibata2014GPGPU} in C++. We use a semi-automatic photogrammetry-based system employing a robot \cite{Mortezapor2022Photogrammabot} to create a precise map of the markers along with an accurate virtual representation of the environment. 

\subsubsection{Head Tracking}
For head tracking we use the IDS camera uEye UI-3251LE with 1.92 megapixel and a fish-eye lens. Although capable of 60 frames per second (fps), we run tracking usually just at 12 fps, because tracking data is fused with Simultaneous Localization and Mapping (SLAM) of the HMDs. Thus, the highly accurate absolute pose is calculated from markers, while the relative optical and inertial tracking from the HMD provides high frame-rate. Central to configuration and processing of tracking data is the Tracking Manager. On the Unity side its functionality is accessible through a number of wrapper classes and can be conveniently configured through the UI. Tracking devices can be easily added, operations (e.g. coordinate space transformation) quickly defined, and mapping to a user’s virtual representation conveniently set. At runtime tracking data is updated and propagated through the different stages in the framework.
The implementation of the tracking devices, data handling, filtering, skeleton transformation etc. is encapsulated in a separate assembly and can be accessed from a simple public interface.

\subsubsection{Object Tracking}

\begin{figure}[tb]
 \centering 
 \includegraphics[width=\columnwidth]{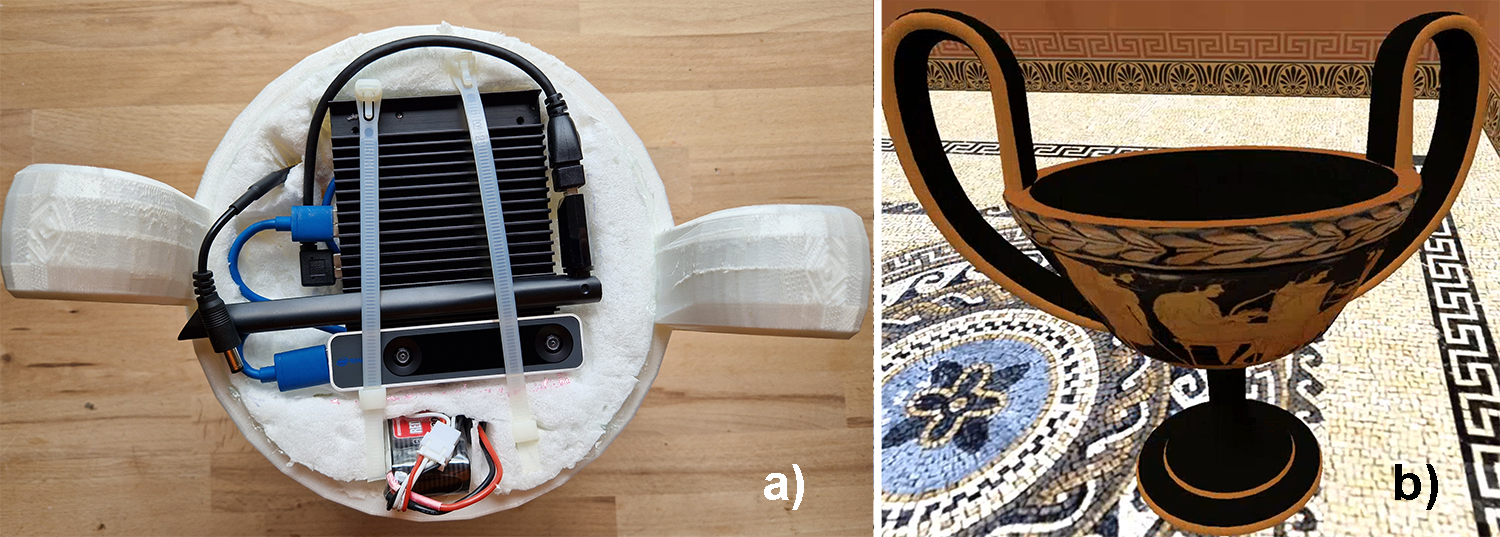}
 \caption{a) Prototype of the Object Tracking Module in a 3D printed vase. b) VR view of the vase in the informal learning experience.}
 \label{fig:Object_Tracking_Vase}
\end{figure}

\begin{figure}[tb]
 \centering 
 \includegraphics[width=\columnwidth]{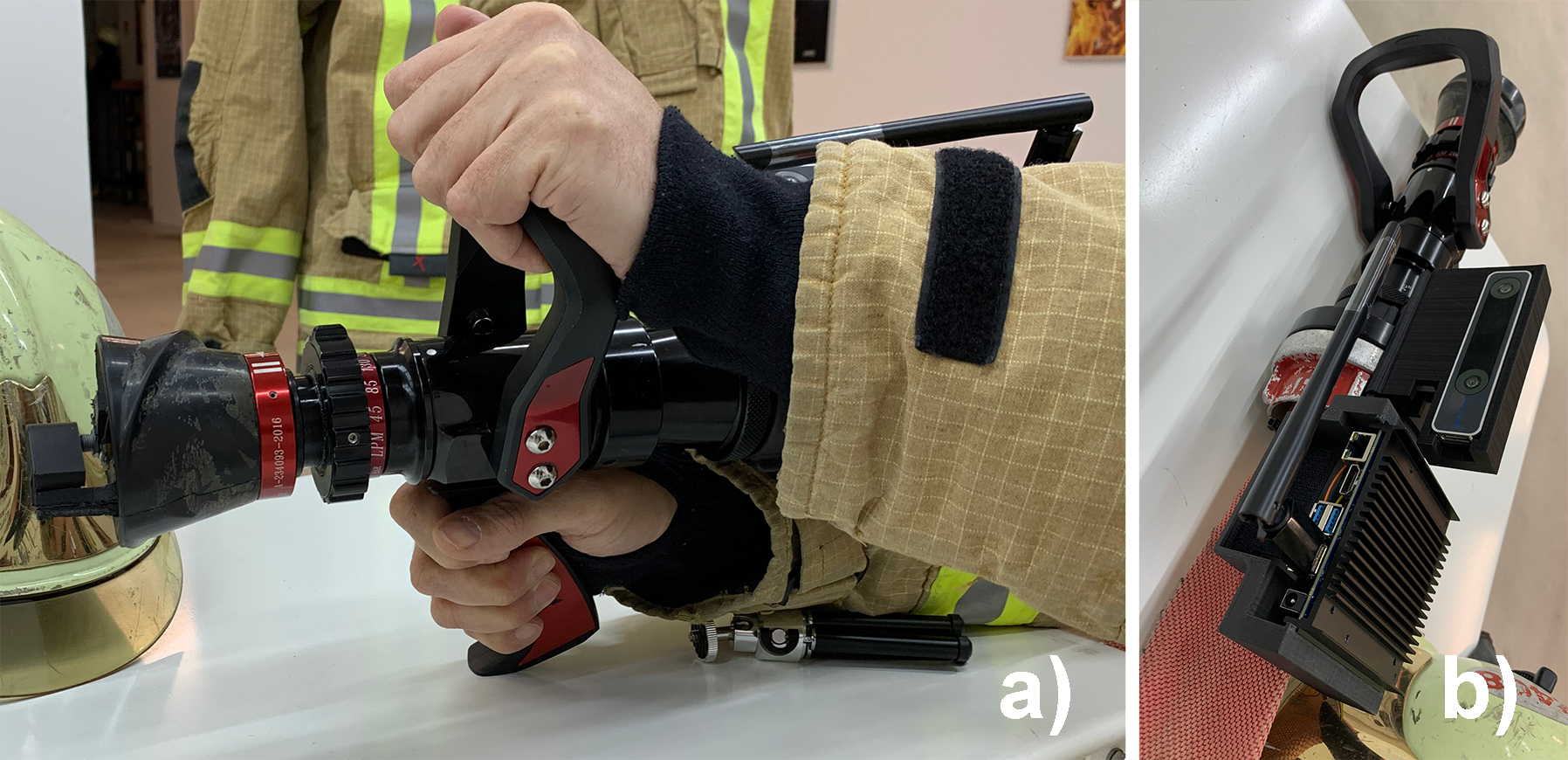}
 \caption{a) Prototype of the Object Tracking Module installed on a fire hose with operational levers from side and b) top view showing RealSense sensor and Odroid computing unit.}
 \label{fig:Object_Tracking_Hose}
\end{figure}

To integrate real physical objects into the virtual environment, thus emphasizing MR in the platform, it provides an object tracking solution. It consists of self-contained inside-out tracking modules with relatively small size of 10 cm side-length. Linux based Odroid N2+ Mini-PCs run the tracking software that is based on the marker-tracking solution used for head-tracking in the ImmersiveDeck. The object tracking module support Intel RealSense T256 sensors via the Intel RealSense SDK 2.0. Tracking data is broadcast to the other devices via WiFi. Finally, custom interaction code allows application of the tracked pose to be applied via Drag-and-Drop to an arbitrary virtual object.
With the marker-tracking implementation the object tracking module already tracks the global pose of the sensor and therefore an object it is attached to. However, due to the limited computational power of a Mini-PC, the framerate of the marker tracking is relatively low and not sufficient for a good MR experience. Therefore, it fuses the global pose from the marker-tracking with the relative pose retrieved from the Intel RealSense sensor, which uses IMU and optical data for SLAM. \autoref{fig:Object_Tracking_Vase} and \autoref{fig:Object_Tracking_Hose} show prototypes and screenshots of the object tracking module with objects of the two respective scenarios. For objects or tools bound to an individual user, a controller can be used for tracking as shown in \autoref{fig:Teaser} a) and b) with the fire hose. Usually, however, it is passed between users and therefore employs the object tracking module.

\subsection{Interaction and Collaboration}
Various modules for intuitive interaction have been implemented in the ImmersiveDeck. Hands and fingers are tracked optically and represented in the virtual world, thus allowing non-verbal communication with gestures. All interactions and their effects are distributed in the multi-user environment. Users can shake hands as they do in real life and navigate the virtual naturally by physically walking around in the real world \cite{Nilsson2018NaturalWalking}. Furthermore, redirected walking techniques have been implemented to allow the users to scale between different floors, while teleportation allows the users to reach positions out of the physical space, if required.



Standard configuration for interactions for 3D objects (being grabbable by the user or attached to the user, opening of drawers, doors, the UI-System providing buttons and touchpad interaction, etc.) are easily applied using the appropriate scripts and template objects provided by the platform.

Furthermore, it is possible to interact with and touch mixed virtual-real objects shown in \autoref{fig:Object_Tracking_Vase} and \autoref{fig:Object_Tracking_Hose}, if the virtual and real representations are registered dynamically or statically as described in the previous subsection.

\subsection{Networking and Multi-User}
Distribution in a multi-user setting is implemented in a basic client-server setup, where a WiFi router (Asus RT-AX88) is connected to the server, while the VR backpacks use their internal WiFi cards.
As described in \cite{podkosova2016ImmersiveDeck} the challenge in a multi-user XR system is to distribute large amounts of tracking data, while keeping the perceived latency low. Especially in mixed virtual-real environments (close to) synchronous perception of visual and haptic feedback is important to not break the immersion, e.g. in situations when users touch each other or a virtual-real object \cite{podkosova2016ImmersiveDeck}.
Due to this trade-off we configured the synchronization of object properties carefully.
 Unity UNet networking is used in the presented ImmersiveDeck implementation. For convenient use we have implemented a networking module for easy configuration of the network, creation of distributed objects, triggering of network-callbacks and automatic distributed scene management.  

To improve networking capabilities for larger numbers of users we optimized data transmission on a packet level. Especially skeleton-movement data generates a large amount of data per user, since multiple joint rotations have to be synchronized with the server and other clients. Therefore, reducing or compressing this data can also substantially decrease the required bandwidth. For one, only noticeable changes are transferred with sufficient accuracy (I.e. no changes below one tenth of a degree). Secondly, the resolution of the transmitted rotation is also reduced to approximately this value.

In addition, the ImmersiveDeck implements player management that provides a number of features in a multi-user environment. Firstly, the player setup can be conveniently handled from the server-side as described earlier. Secondly, player connection and reconnection is automatically handled by the ImmersiveDeck. Finally, different player types can be easily configured.

\subsection{Multi-sensorial Stimuli}
Beside (passive) haptics as described in \autoref{subsec:ID:TrackingAndMapping} the ImmersiveDeck provides smell, wind and heat beyond the standard video and audio output modalities. Therefore, the ImmersiveDeck integrates the Olorama device (smell) as well as a wireless network interface that can control a ventilator and infrared heater. 

\section{Informal Learning}
\begin{figure}[tb]
 \centering 
 \includegraphics[width=\columnwidth]{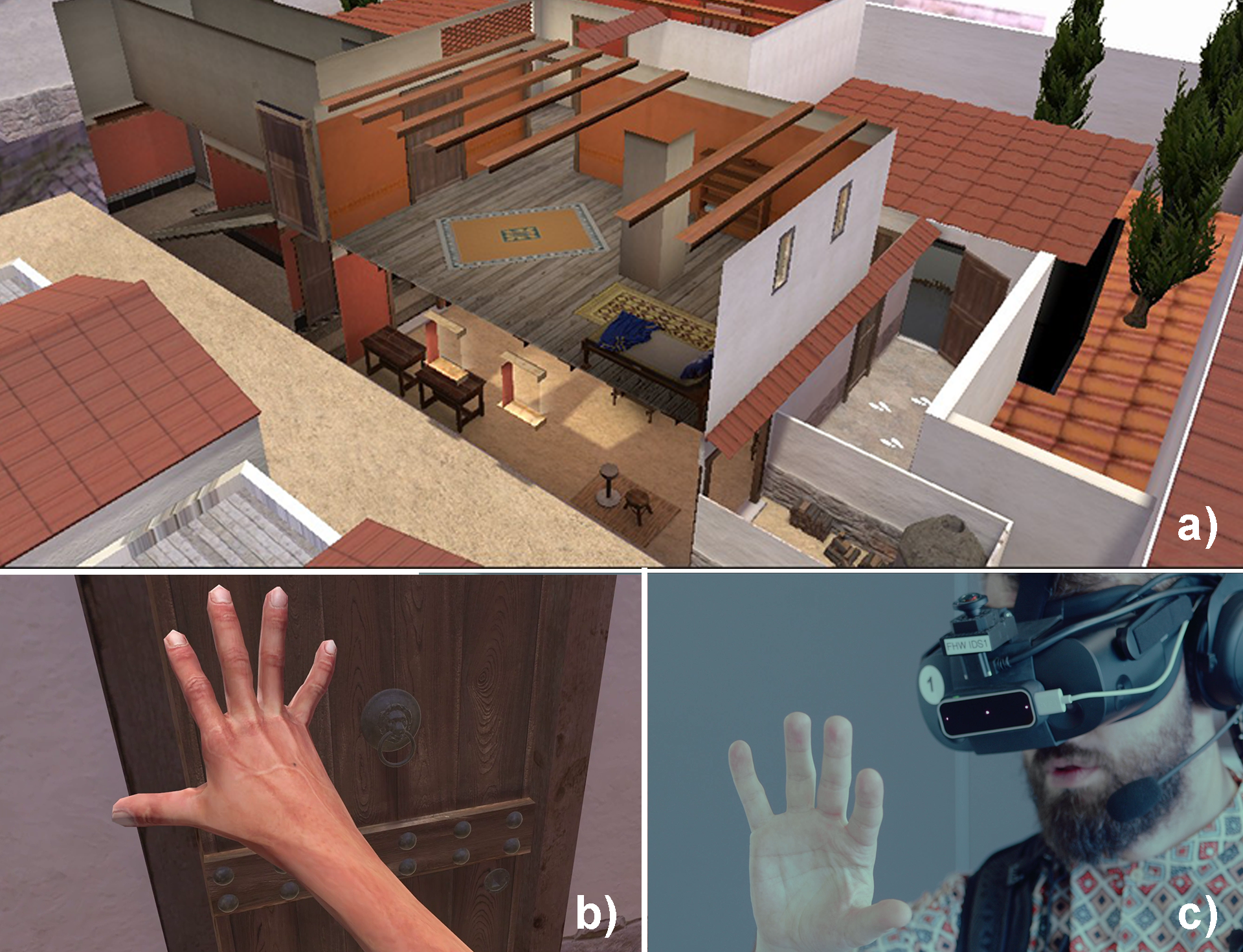}
 \caption{Informal learning experience. a) Partially transparent view of the two-storey Athenian house with oboarding area (right), pottery workshop (lower middle), dressing room (upper middle), staircase (left). b) Avatar's animated hand. c) Photo of a user with ImmersiveDeck setup including Ultraleap for hand tracking.}
 \label{fig:Athenian_House}
\end{figure}

The scenario design for an informal learning experience was created together with the Foundation of the Hellenic World and an Advisory Board of historians and archaeologists, including experts from the Museum of Cycladic Art in Athens. Scenario design and development was carried out in a participatory and iterative manner, through a series of onsite workshops and body-storming sessions involving all stakeholders and eliciting their requirements using a user-centered design approach \cite{roussou2021requirements}.
The result is the informal learning scenario ``Ergochare’s house: an eventful night in an Athenian house of the 5th century BCE'' (Athenian house, for short), which combines historical accuracy and dramatic storytelling to enhance knowledge about the past and historical empathy. The experience is set in an Athenian home in Classical Athens during the Peloponnesian War. It allows the users to partake in and learn about the everyday life of people and how they lived and participated in the different events of the community. Visitors can take on the roles of the members of an Athenian family in their house (\autoref{fig:Athenian_House}) and can among other things partake in a farewell celebration of a soldier and in crafting activities in a pottery shop (\autoref{fig:Teaser} b)). In the interactive guided storytelling experience the users can roam by natural walking, communicate and interact freely, while occasionally being prompted by non-player characters for certain actions to progress the story-line, which is implemented in various game-states. A voting mechanism built on the UI-system and additional states allows branching of the story-line. The visitors start in an onboarding area outside the virtual house being able to become accustomed to the technology as well as receiving instructions on the experience. The experience provides several opportunities to interact with grabbable virtual and Mixed Reality objects (e.g. as illustrated in \autoref{fig:Teaser} and \autoref{fig:Object_Tracking_Vase}, \autoref{fig:Athenian_House}), while informing about the cultural and historical background. The ImmersiveDeck's redirected walking allows access to the second floor of the house.

Such an interactive, social and participatory multi sensorial experience, combining historical accuracy in representation with dramatic storytelling, can enhance knowledge about the past, historical empathy, and resonance \cite{perry2017engaging,roussou2021requirements}.

While groups of 10 people are desirable in this context of informal learning, the experience has been designed for five users simultaneously. Nevertheless, we tested whether the platform would scale up to 10 users, as described in \autoref{subsec:TechnicalAssessment:Networking}.

\section{First Responder Training}
\begin{figure}[tb]
 \centering 
 \includegraphics[width=\columnwidth]{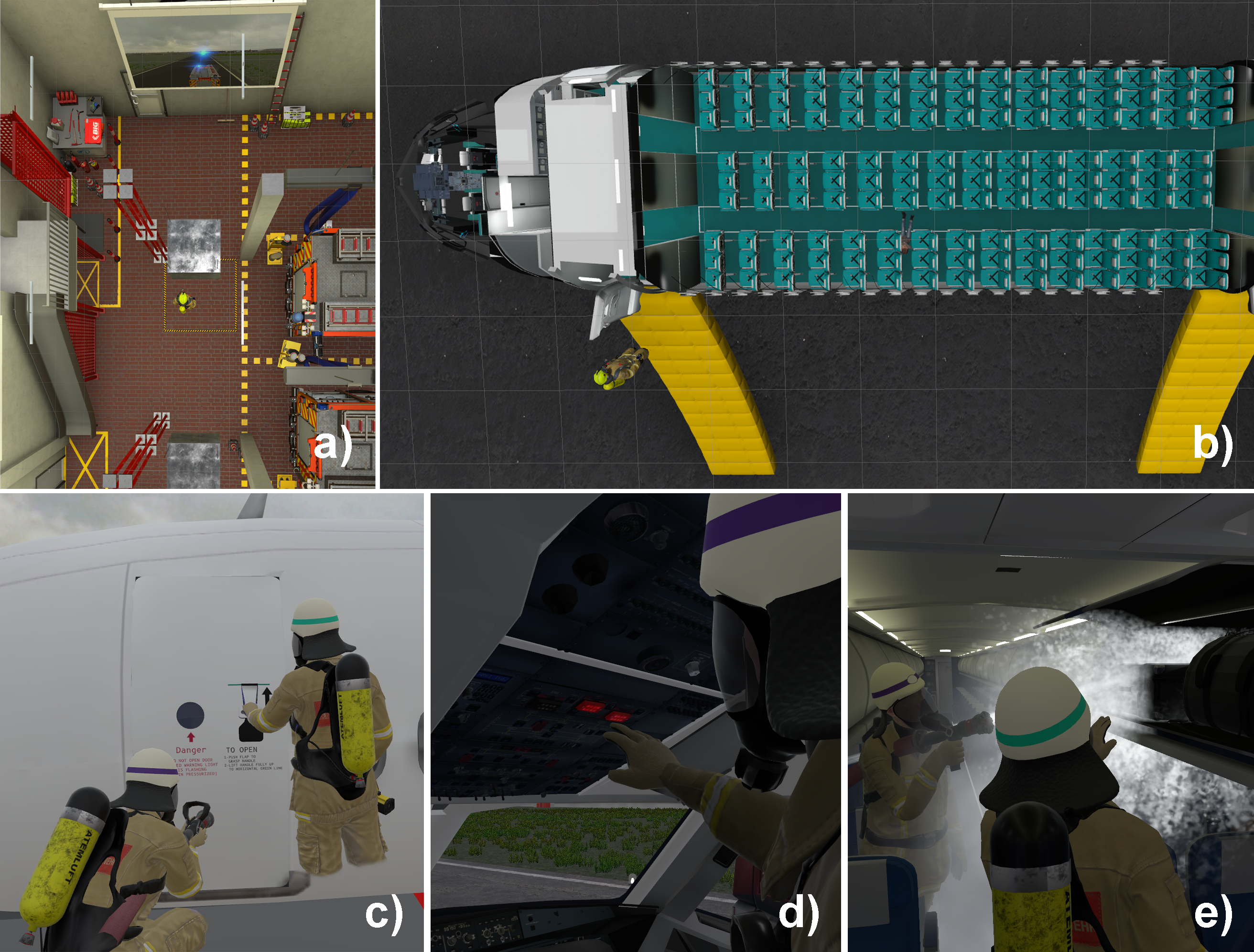}
 \caption{Birdseye view of the main scenes of ARFF a) fire station garage with onboarding and practice area and briefing screen (top) and b) plane interior and exterior training area. c) Firefighters opening the plane's door d) shutting down possibly dangerous systems in cockpit and e) extinguishing cabin fire in overhead compartment.}
 \label{fig:ARFF_Scenes}
\end{figure}

\begin{figure}[tb]
 \centering 
 \includegraphics[width=\columnwidth]{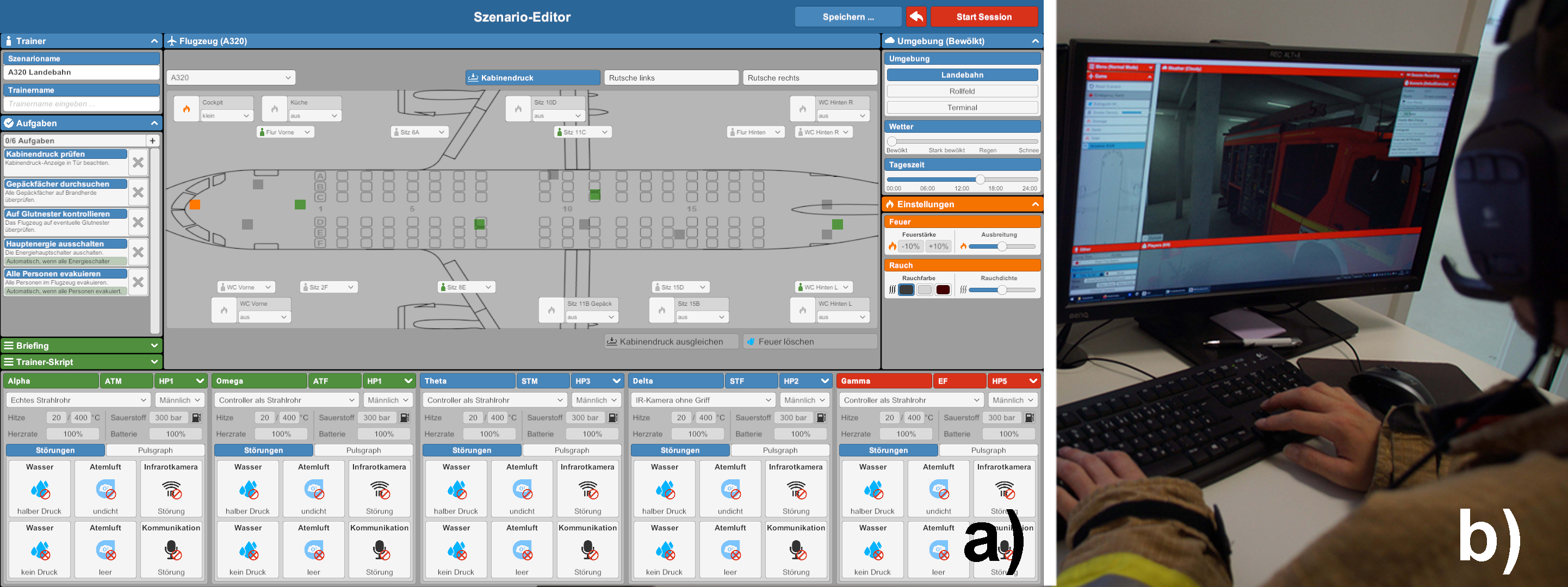}
 \caption{ARFF scenario showing a) scenario editor and b) trainer following the simulation in the administrator interface.}
 \label{fig:ARFF_Szenario_Editor_Operator}
\end{figure}

Based on the ImmersiveDeck a first responder training experience called ``Aircraft Rescue and Fire Fighting XR Training (ARFF)'' has been created, which simulates an emergency situation with a plane at an airport. It has been developed together with and following requirements of two professional airport fire brigades (of Flughafen Berlin Brandenburg and Athens International Airport). The experience was created in a user-centered approach based on the analysis of the firefighters' didactic objectives and goals together with the techniques introduced in \cite{roussou2021requirements} and described in the previous section.
The simulation allows the first response team to evaluate the situation and coordinate further actions as part of a standard procedure. The main goals of such a training are reconnaissance of the situation, communication with HQ, team members, and passengers and the habituation to stress, views of such scenery, the sounds as well as the repetition of standard procedures. An Airbus A320 and a Boeing 767 (\autoref{fig:ARFF_Scenes} b)) airplane with interior and exterior have been modelled and integrated as a training environment alongside with exterior parts of the airport and interior of the fire-station as starting point of the training experience (\autoref{fig:ARFF_Scenes} a)). Many objects and vehicles to populate the scenario have been modelled or sourced from asset stores or partner studios as can be seen in \autoref{fig:ARFF_Scenes}. 
In preparation for a training session, the trainer can use the scenario editor shown in \autoref{fig:ARFF_Szenario_Editor_Operator} a) to choose level of fire, density of smoke, amount, type and place of passengers and injured persons. Levels of difficulty can be adjusted individually for different teams and their skills. The trainer observes the simulation via a monitor and a “virtual” radio connection (\autoref{fig:ARFF_Szenario_Editor_Operator} b)), can interfere and trigger special occurrences or even record and remark specific situations for later debriefing. 

In such an indicative scenario, usually two to five members of this first response team will enter the training scenario together. The first response team starts in the fire station at the airport (\autoref{fig:ARFF_Scenes} a)). There users can get accustomed to the virtual environment and interaction possibilities. After that the trainer can sound the alarm and display a video showing the approach to the emergency site and providing briefing information. Finally, the team deploys to the emergency site in front of the airplane, where it can choose the entry.

The team has to open the door in the correct procedure in order to get inside the cabin as shown in \autoref{fig:ARFF_Scenes} c). Smoke and obstacles make it difficult to find the way through the aisle. The primary task is to find all injured persons, secure them and classify their status. Furthermore, trainees have to apprehend the changing situation and detect additional threats. Throughout the duration of the XR experience, the status, location, and observations of the team have to be communicated by radio to the operation control. The team has to clear the situation on the inside (evacuate people and extinguish fires) as illustrated in \autoref{fig:ARFF_Scenes} d) and e), while the fire trucks contain the fire on the outside. Once the mission is accomplished, the team can head back to engage in a debriefing in which the entire experience, that was recorded for playback, can be reviewed and discussed among the team and their trainers.



\section{Technical Assessment}
\subsection{Object Tracking}
\begin{figure}[tb]
 \centering 
 \includegraphics[width=\columnwidth]{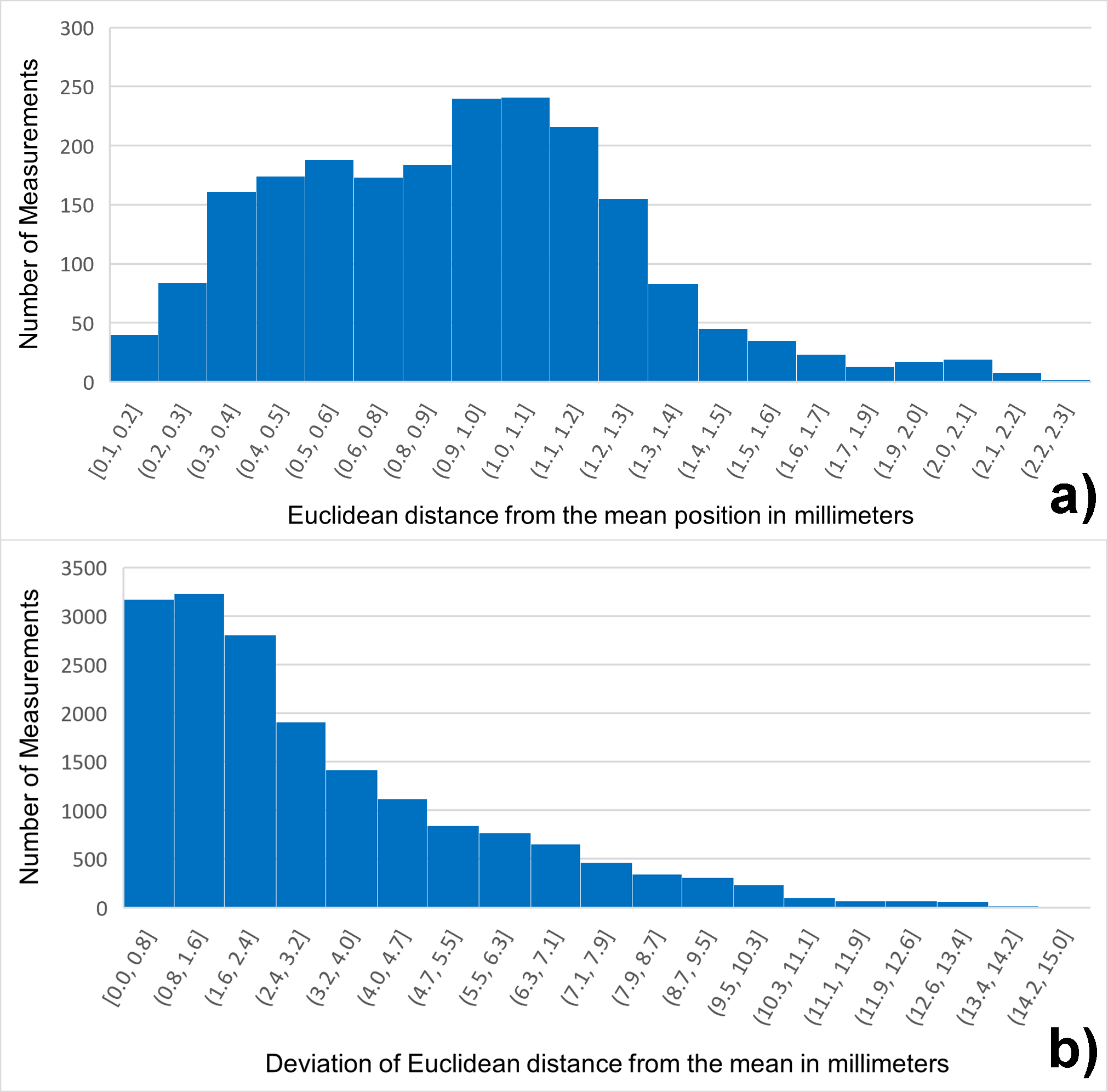}
 \caption{Distance measurements of a tracked object (e.g. fire hose) showing a) static jitter and b) relative accuracy during walking.}
 \label{fig:Jitter_Accuracy}
\end{figure}

We have tested, if the quality of object tracking is sufficient for the two XR experiences. The objects are mostly used in close proximity to the users, which is why it is important that jitter is small (e.g. for aiming the fire hose) and relative accuracy is good (for handing over objects). 
We have measured static jitter by placing a single object tracking module in the middle of the mapped space and recorded the position over 2100 frames. The root-mean-square distance of these positions from the mean center position is 0.96 mm. 
\autoref{fig:Jitter_Accuracy} a) shows the histogram of the distances.
To assess the accuracy of the tracked position we followed a similar approach to \cite{mossel2021immersive,podkosova2016ImmersiveDeck}. Two sensors were mounted at fixed distance to each other on a metal bar, with the cameras facing upward. We moved that construction through the mapped space at walking speed for over 17500 frames at about a meter height, where we would assume many interactions would take place. The root-mean-square distance from the mean distance was measured with 4.05 mm, while a histogram of the distances is illustrated in \autoref{fig:Jitter_Accuracy} b). 

Finally, we moved the metal bar through the mapped space quickly for over 21000 frames, twisting and turning it in the process, similar to wielding a sword. We consider this the worst case, where sensors would point at the markers only occasionally, which is not very likely to happen in the described scenarios. Here, the root-mean-square error increased to 14.16 mm. 

\subsection{Networking}
\label{subsec:TechnicalAssessment:Networking}
For the ARFF scenario it can be desirable to have additional users in the system (e.g. for operation of fire trucks, backup team members, spectators). Also for Athenian house more users in the experience would increase throughput, which is important for larger groups of visitors like school classes. Therefore, we implemented a toolset based on the ImmersiveDeck and assessed feasibility of 10 simultaneous users in the experiences. The toolset consists of multiple hard- and software components to systematically assess network performance and run simulations to identify latency issues and network related bottlenecks (e.g. through bandwidth limitations).

The hardware consists of 11 PCs connected via a WiFi router. 
On the software side, we have implemented tools in Unity based on the ImmersiveDeck. The main challenge for a sizeable number of users lies in the synchronization of all users' movements over all clients and the server, because of the large amount of transmitted data (many joints in the skeleton) and requirement of low latency for interaction. The toolset has been designed to measure multiple parameters. The main focus lies on latency, but also calculation times of certain critical processing steps and update rates / fps  and transmitted data are being assessed. Due to the spatial constraints in this paper, here we will focus on certain latency measurements, while the other measurements provide similar indications.
The test application is graphically very simple to avoid unnecessary overhead from rendering. The scene only contains the users’ avatars respectively their skeletons. 
To create a realistic load on the network we have implemented recorder/-player scripts for human movements. 

The main latency measurements are taken during test series playing back these skeleton movements. The most important measurement is the time delay of two similar poses being available for processing between a locally played back skeleton recording and the same recording played back on a remote client. We refer to this measurement as ``pose latency'' in the following.

To test latency, we ran test series of replayed skeleton movements with various client and parameter configurations. For most tests we used several recordings of around 10 seconds
. We used recordings with different amounts of movement, since our data compression algorithm compresses better when there is little change from frame to frame. However, differences between the movements have shown to be relatively small, which is why we only present results from the  ``dance''-movement here, where hands and head are moved quickly.

\begin{figure}[tb]
 \centering 
 \includegraphics[width=\columnwidth]{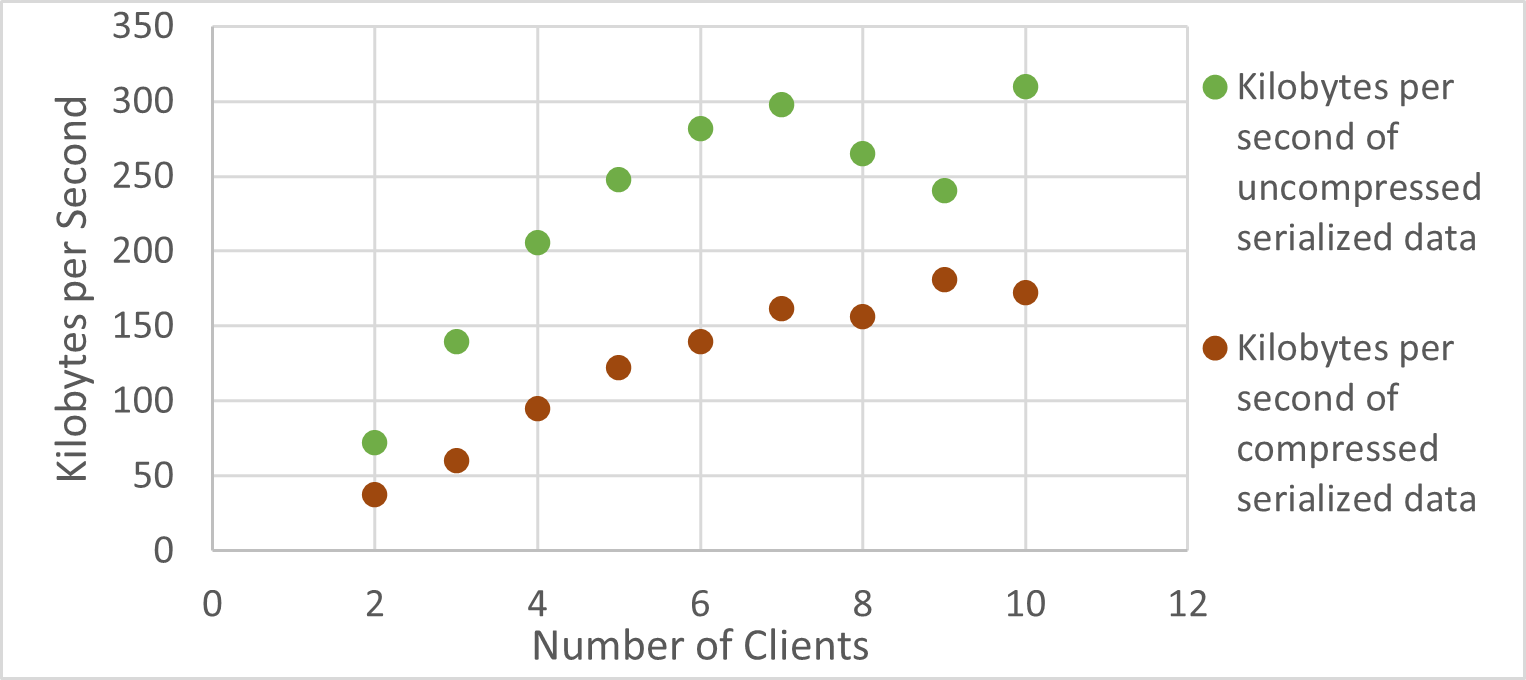}
 \caption{Amount of serialized pose data.}
 \label{fig:Serialized_Data}
\end{figure}

There is a significant difference between the amount of pose data being (de-)serialized with and without compression as can be seen in \autoref{fig:Serialized_Data}. Without compression the amount of data increases by approximately a factor of 2.3, i.e from 5.2 kilobyte per second to 11.9 kilobyte per second for the dance sequence. 

\begin{figure}[tb]
 \centering 
 \includegraphics[width=\columnwidth]{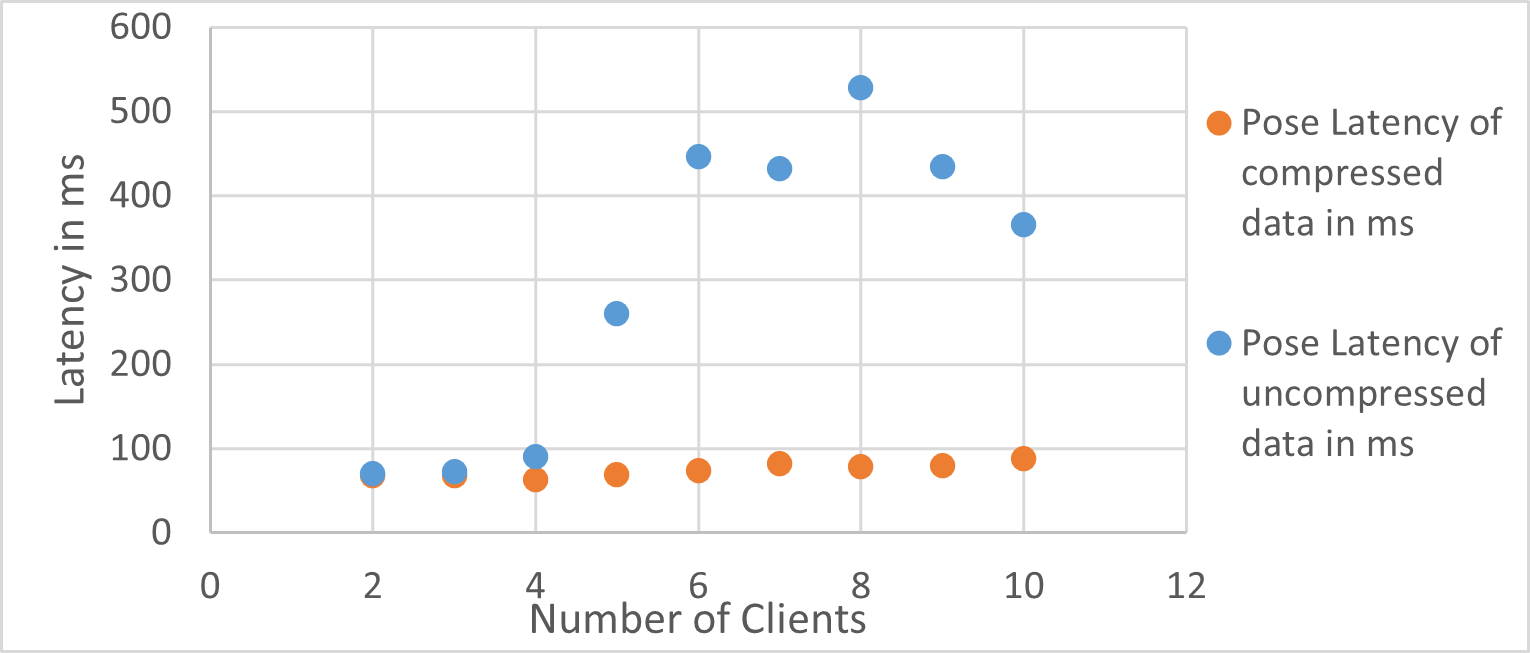}
 \caption{Pose latency measurements over number of clients.}
 \label{fig:Pose_Latency}
\end{figure}

Without compression, distribution of skeleton pose data worked fine for 4-5 users, but with more than those latencies would quickly increase to a degree where multiuser-interaction is not possible anymore. Latencies in that case go from well below 100ms up to and over 500ms. \autoref{fig:Pose_Latency} illustrates data from the dance sequence updated at 10ms intervals. With the larger send intervals usually employed, the limit might be hit at a later client number as long as only skeleton data is transmitted. However, with most MR-applications also object poses and other data has to be distributed between clients and server. Therefore, the somewhat simplified illustration is not an unrealistic one. From this and other test cases we can observe that somewhere between 200 and 350 kilobytes per second (probably depending on packet size) is a limit of data that can be serialized and transmitted with the used netcode implementation and network hardware. Over that limit network packets will get dropped and data not processed.

Even with a larger number of users the latencies stay relatively stable, when compression is used. Only with higher send-rates (i.e. at 10 ms intervals) an increase in latency can be observed, as illustrated in \autoref{fig:Pose_Latency}. Pose latency increases from around 65 ms for lower one-figure client numbers to around 85 ms for 10 clients. Furthermore, as expected, the amount of (de)serialized data increases with the number of clients. However, while from 2 to 7 clients the increase is rather linear, with 8-10 clients the amount of data stagnates, which again leads to the assumption that not all the data can be transmitted/processed anymore. Note that this is only for high send-rates, while for the usual send-rate (with 100 ms intervals) the influence of the number of clients is less significant on the latency and the serialized data increases as expected.



\section{Preliminary Expert Assessment}
To evaluate the user experience of content creation for and with the ImmersiveDeck, we conducted a preliminary informal evaluation following Lam et al. \cite{lam2012Empirical}.
At this stage two teams of experts have been working with the platform creating the respective two experiences for informal learning and first responder training. However, the team working on first responder training strongly overlaps with the platform developers. Therefore, to avoid bias, our assessment is focussed stronger on the informal learning experience. We evaluated content creation with two experts, aged 32 and 53 years, each having over six years of experience in XR content creation (as developer respectively project manager) and more with 3D desktop applications. Domain experts can provide important insight \cite{tadeja2021exploring}, although we advise caution regarding these preliminary results, and recommend a larger number of participants for future work.
The experts worked with the ImmersiveDeck platform for over a year in a joint research project with regular support-meetings from the platform developers until finalization of the experience. After that time period we had a semi-structured discussion with them to evaluate the different aspects of experience creation with the ImmersiveDeck. 

The experts found it easy to integrate their custom made avatars, also due to the ``great'' documentation of that feature. With their readily rigged character they reported to have just needed ``some clicks inside the ImmersiveDeck''.
The state logic, was considered to be the most challenging part by the experts in the development process with the ImmersiveDeck.
One reason for that was, that the state management became relatively complex, because it followed the non-linear story-line of the experience. Therefore, it required substantial consideration regarding logic and parameters of the states.
Another reason was, that the requirement of using state management became apparent relatively late during the development. 

To the experts the platform's tracking and mapping functionality left the biggest impression - positively as well as negatively. Having developed the first prototype of the experience without access to an ImmersiveDeck installation, both mentioned on having it found ``very impressive'' that their prototype would run out-of-the-box, when deployed and allowed navigation by physical walking and other interactions. However, over time it showed necessary to visit the location of installation approximately an average of 2-3 times per month 
to test especially mixed real-virtual interactions. Testing multi-user functionality and configuration like tuning parameters of voice communication has been the second factor besides MR.

The experts liked most of the platform functionalities like grabbing mechanics, the UI interaction and other that exist as template objects and ``with some tweaks and adjustments, they worked just out-of-the-box''. They also considered functionalities related to interaction a ``great feature of the platform'' for content creators, because it drives the immersion and interaction with the environment feels realistic. Furthermore, the functionality of using the real physical environment to create a ``very immersive virtual environment'' has been considered one of the ``highlights''.
Only the grabbing mechanism has been reimplemented by the platform developers following the feedback from the experts during development. First it used a physics based approach, where every finger was checked for collision with the grabbed object. However, in first tests this showed to be not reliable enough. Therefore, now a robust gesture based approach is used, where closing of the hand attaches an object in close proximity to the hand and opening releases it. 

\section{Discussion and Guidelines}
Regarding the technical assessment of the object tracking module, especially static jitter and relative accuracy with slower movements showed relatively small jitter respectively error values in the small one-figured millimeter domain and below. This should very well allow precise interaction, such as aiming of the fire hose or handing over of the vase between users in the experiences, especially since we expect these interactions to happen relatively slow. Even in fast movements the root-mean-square error increased to only approximately 14 mm, which we consider good for situations where probably also the user moves fast and will hardly notice the deviation. In rare cases we observed outliers about one order of magnitude larger than the root-mean-square error. These occur probably in situations where the optical sensor detects no or just blurred markers and features and the pose calculated from inertial data only drifts quickly with fast movements. However, once the sensor detects markers, the pose can be corrected quickly, thus resolving the situation in the context of the proposed experiences.

While for a user's head-tracking a latency of below 15ms is desirable to avoid cybersickness \cite{Elbamby2018Latency}, higher latencies of other users' movements are tolerable until the immersion breaks. For slower interactions, such as shaking hands or handing over of objects as expected in the proposed experiences, 85ms measured for 10 clients should be acceptable.
Thus, the current version of the ImmersiveDeck provides the means to run experiences with up to 10 users.

Here, we briefly discuss and try to derive some guidelines for creating learning and training experiences in an immersive multi-user XR environment from the expert assessment and lessons learned. The experts found many of the ImmersiveDeck's features to work ``out-of-the-box'' and successfully created the experience. Some of their feedback during development has been already implemented, more recent findings from the previous section are set for future work or might serve as general recommendation for content creators:

\textit{Weigh carefully between robust hand (grabbing) interaction and physically correct ones.} With our original physics based approach, robustness of grabbing depended very much on the shape of the object, was problematic in case of occlusions of some fingers and has been re-implemented according to the experts' suggestions.

\textit{Plan regular tests with the actual platform when working on a mixed real-virtual multi-user environment}, especially for MR interaction. Furthermore, for the ImmersiveDeck we are working on additional functionalities to facilitate alignment of the real environment with virtual content and simulation of the physical world for off-site tests following the experts' feedback.

\textit{Management of the state of the experience in the multi-user environment should be considered from the beginning of development}, because adding proper state management later on can be tedious and time-consuming, as reported by our expert content creators. In addition, more elaborate state management by the platform, especially regarding branching story-lines, could facilitate content creation.

\textit{Test the equipment early on in the intended location, especially if you have no or little control over the physical environment.} We have successfully set up the ImmersiveDeck platform in five locations, three of which were not purely dedicated to running XR experiences (which probably only a few end-users of learning/training applications can afford). All three of the latter had their specific challenges. Firstly, reflective and partially transparent glass walls, or uneven lighting challenged optical sensors, especially during mapping, as well as the photogrammetric reconstruction. Secondly, pipes and other structures near the ceiling, produced many occlusions in the photos for the location in \autoref{fig:Reconstr_Firestation_Garage}. That led us to improve our mapping algorithms and together with increasing the number of photos, markers or marker-sizes, all locations were mapped successfully. 



\section{Conclusion and Future Work}
This paper describes the creation of two XR experiences in the context of informal learning and first responder training based on the ImmersiveDeck platform. Furthermore, it illustrates the design and implementation of the platform features required in the respective learning and training domains. We evaluated platform technologies with a view towards the requirements of the scenarios and showed applicability. 
Finally, we conducted an informal expert evaluation of the content creation process and present derived guidelines.

We are in the process of conducting extensive and thorough user studies to validate the platform and the created training/learning XR experiences in the real world, i.e., at two major international airports in Germany and Greece, and at a highly visited informal educational and cultural center in Greece.

\acknowledgments{
This work has been funded by the European Union’s Horizon 2020 research and innovation programme within the BRIDGES project under grant No 952043. We thank all partners for their contribution.
}

\bibliographystyle{abbrv-doi}

\bibliography{library}
\end{document}